\documentclass[aip, reprint, superscriptaddress,amsfonts,amssymb,amsmath]{revtex4-1}
\usepackage{graphicx,xcolor}
\usepackage{epstopdf,epsfig}

\newcommand*\diff{\mathop{}\!\mathrm{d}}
\usepackage[normalem]{ulem}
\usepackage{bm}
\usepackage{relsize}
\allowdisplaybreaks
\usepackage[percent]{overpic}

\newcommand{\bu}{\mathbf{u}}
\renewcommand{\Re}{\text{Re}}
\newcommand{\Wi}{\text{Wi}}
\makeatletter
\def\blfootnote{\xdef\@thefnmark{}\@footnotetext}
\makeatother
\begin{document}


\title{Viscoelastic laminar drag bounds in pipe flow}


\author{M. Malik}
\email[]{dr.malik.barak@gmail.com}
\author{Roland Bouffanais}
\email[]{bouffanais@sutd.edu.sg}
\affiliation{Singapore University of Technology and Design, 8 Somapah Road, Singapore 487372}
\author{Martin Skote}
\email[]{m.skote@cranfield.ac.uk}
\affiliation{Cranfield University, UK}

\begin{abstract}
The velocity and friction properties of laminar pipe flow of a viscoelastic solution are bounded by the corresponding values for two Newtonian fluids, namely, the solvent and a fluid with a viscosity identical to the total viscosity of the solution. The lower friction factor for the flow of the solution when compared to the latter is tracked to an increased strain rate needed to enhance viscous dissipation. Lastly, we show analytically that the effective viscosity varies similarly to the radial diagonal component of the conformation tensor as observed numerically in turbulent flows, and give a lucid interpretation of shear-thinning through a sequence of underlying constitutive physical phenomena.
\end{abstract}
\maketitle
Elasticity\blfootnote{\hspace{-15pt} This letter accepted by Physics of Fluids could be found after publication at https://aip.scitation.org/journal/phf} effects have long been known to affect transition to turbulence and drag in fluid flows, whether these effects are induced by fluid-structure interaction~\cite{dixon1994optimization,carpenter2001progress, malik2018growth}, or by viscoelastic rheological aspects associated with non-Newtonian fluids~\citep{toms1948some, sureshkumar1995linear,zhang2013linear, garg2018viscoelastic, brandi2019dns}. For instance, the addition of low concentrations of long-chain polymers generates an astounding 80\% drag reduction (DR) in turbulent regimes~\citep{white2008mechanics, graham2004drag}, which has significant implications for practical applications. At relatively low Reynolds numbers, \emph{elasto-inertial} effects sets in~\cite{samanta2013elasto, dubief2013mechanism}, which are also observed in the instabilities of flow bounded by compliant walls~\cite{dixon1994optimization,carpenter2001progress, malik2018growth}.

When considering turbulent dilute polymeric solutions, the exhibited DR has been attributed to the interplay between flow turbulence and elasticity of the polymers in the near-wall region~\cite{lumley1969drag,nieuwstadt2001drag,graham2004drag, white2008mechanics}. DR is most prominent when the time scale of the polymer elastic dynamics---known to be dependent on the number and length of monomers making the polymer---is of the same order or higher than that of the turbulent fluid flow~\cite{sreenivasan2000onset}. In addition, such fluids exhibit a maximum drag asymptote with respect to the polymer concentration~\cite{virk1967toms,kenis1971friction,sreenivasan2000onset, benzi2006maximum, xi2012dynamics, lopez2019dynamics}, which is suggested to be associated with elasto-inertial instability~\cite{samanta2013elasto, dubief2013mechanism}. A phenomenon of reverse transition has even been uncovered in such flow recently~\cite{choueiri2018exceeding,pereira2019beyond}. The energy cascade is also different from that of the Newtonian counterpart~\cite{perlekar2006manifestations} since some energy is rerouted to the polymer stretching dynamics, thereby reducing the formation of the smallest eddies and the associated viscous energy dissipation.

However, this DR phenomenology is absent in steady laminar flows with polymers~\cite{virk1967toms, white2008mechanics} due to absence of small time scale in the flow dynamics. Here, we show analytically that the laminar drag of a FENE-P fluid (the solution) in a cylindrical pipe exhibits a set of lower and upper bounds. Specifically, the laminar drag is lower than that of a Newtonian fluid with a viscosity matching that of the total viscosity of the solution, while being higher than that of the pure solvent. This previously unreported phenomenon is due to the effective viscosity of the solution being bounded by the limits of that of two Newtonian flows: (i) the flow of pure solvent and (ii) that of a viscosity-matched fluid. 

Finally, we relate the radial component of the stress tensor to the effective viscosity, which has observational support from direct numerical simulations of turbulent flows~\cite{benzi2010short}. Based on this result, we give an interpretation of shear-thinning, which is well-known to be directly correlated with the axial elongation of polymer molecules. We argue that the perceived effect of shear-thinning is due to a sequence of constituent fundamental physical phenomena, such as elasticity and force balancing wrapped together.

The flow dynamics is characterized by three nondimensional parameters: (1) $\Re=\rho U_cR/\mu$, (2) $\beta=\mu_s/\mu$, and (3) $\Wi=\lambda U_c/R$, with $\mu$ (resp. $\mu_s$) the total viscosity (resp. the solvent viscosity),  $U_c$ the centerline velocity in the absence of polymer, $R$ the pipe radius, and $\lambda$ the elastic relaxation time. A steady flow of dilute polymers modeled as dumbbells is governed by~\citep[see for example,][]{phan2017understanding} 
\begin{eqnarray}
&\bu  \cdot \boldsymbol{\nabla}\bu = -\boldsymbol{\nabla} p + {\Re}^{-1} \left [{\beta}\boldsymbol{\nabla}^2\bu 
+ {(1-\beta)}\boldsymbol{\nabla}\cdot\boldsymbol{\tau}\right ],\label{gov_eqn_u} \\
&\bu \cdot \boldsymbol{\nabla} \textbf{\textit{\textsf{c}}} - 
\textbf{\textit{\textsf{c}}}\cdot\boldsymbol{\nabla}\bu 
-(\boldsymbol{\nabla}\bu)^{\text{T}}\cdot\textbf{\textit{\textsf{c}}} = -\boldsymbol{\tau}, \label{gov_eqn_c} 
\end{eqnarray}
where $\boldsymbol{\tau} = (f\textbf{\textit{\textsf{c}}}-\textbf{\textit{\textsf{I}}})/\Wi$ is the elastic stress of the polymers of maximum extensibility $L$. Under the FENE-P model, the Peterlin function takes the form $f = (L^2-3)/(L^2-\text{tr}(\textbf{\textit{\textsf{c}}}))$. The conformation tensor has for entries $c_{ij}=\langle \tilde{R}_i \tilde{R}_j \rangle$, where $\tilde{R}_i$ is the end-to-end vector of a polymer molecule, and $\textbf{\textit{\textsf{I}}}$ denotes the identity matrix. For a formulation of FENE-P model alternative to Eqs.~(\ref{gov_eqn_u}) and~(\ref{gov_eqn_c}) with an extended set of parameters and a different set of unknowns, see, for example, Bird \emph{et al.}~\cite{bird2007transport}, which has been solved by Cruz \emph{et al.}~\cite{cruz2005analytical} for laminar profiles in pipe and channel. (For the profiles for inviscid solvent, see Oliveira~\cite{oliveira2002exact}). However, the FENE-P  model is widely known to the community studying turbulence and transition as it appear in Eqs.~(\ref{gov_eqn_u}) and~(\ref{gov_eqn_c}) with the Peterlin function $f$ as mentioned. We therefore solve these equations below and use the resulting analytic solutions to study the bounds for velocity and friction. The laminar profiles derived here will also enable the study the transition to turbulence by perturbing Eqs.~(\ref{gov_eqn_u}) and~(\ref{gov_eqn_c}). Hitherto, the absence of a solution to Eqs.~(\ref{gov_eqn_u}) and~(\ref{gov_eqn_c}) has resulted in that transition studies have only been conducted under the Oldroyd-B model~\cite{garg2018viscoelastic}.

For steady laminar flows, $p =P(x)$, $\textbf{\textit{\textsf{c}}} = \textbf{\textit{\textsf{C}}}(r)$, and $\boldsymbol{u}= [U(r), 0, 0]$ using cylindrical coordinates $(x,r,\theta)$, which are in axial ($\boldsymbol{\hat{e}_1}$), radial ($\boldsymbol{\hat{e}_2}$) and azimuthal ($\boldsymbol{\hat{e}_3}$) directions, respectively. The components of $\boldsymbol{\nabla} \cdot \boldsymbol{\tau}$ in Eq.~\eqref{gov_eqn_u} can be found in Bird \emph{et al.}~\cite{bird1987dynamics}, while Eq.~\eqref{gov_eqn_c} becomes 
\begin{align}
U_r&\left (  
\begin{array}{lcr}
2C_{12} & C_{22}  & C_{23}\\
C_{22} & 0 & 0\\
C_{23} &  0 & 0
\end{array}
\right )
\nonumber \\ 
&= 
\frac{1}{\Wi}\left ( 
\begin{array}{lcr}
FC_{11}-1 & FC_{12}  & FC_{13}\\
FC_{12} & FC_{22}-1 & FC_{23}\\
FC_{13} &  FC_{23} & FC_{33}-1
\end{array}
\right ),\label{C_eqn}
\end{align}
where $F(r)=(L^2-3)/(L^2-\text{tr}(\textbf{\textit{\textsf{C}}}))$ and the subscript $r$ denotes derivative along the radial direction. Equation~\eqref{C_eqn} gives the non-zero elements of~$\textbf{\textit{\textsf{C}}}$:
\begin{align}
C_{11} &= (2\Wi^2U_r^2+F^2)/F^3, \label{eq_C11} \\
C_{22} &= C_{33}= 1/F,\label{eq_C22} \\
C_{12} &= \Wi \,U_r/F^2.  \label{eq_C12} 
\end{align}
Equation~\eqref{eq_C22} implies that the transverse components of $\boldsymbol{\tau}$ follows $\tau_{22}=\tau_{33}=0$, which is in agreement with Ref.~\cite{cruz2005analytical}. Substituting Eqs.~(\ref{eq_C11})--(\ref{eq_C12}) in $F$ yields:
\begin{equation}
F^2(F-1) = 2\Wi^2U_r^2/L^2. \label{f3f2eqn}
\end{equation}
Since the driving pressure gradient in a pipe is given by 
\begin{equation}
\diff P/\diff x = -4/\Re,\label{pressure-gradient}
\end{equation}
 Eq.~\eqref{gov_eqn_u} can be written as, 
$
\beta \left ( rU_r \right )_r + (1-\beta)\left ( {rU_r}/{F} \right )_r = -4r,
$
which under the condition $U_r(0) = 0$ gives
\begin{equation}
U_r = -2rF[\beta(F-1)+1]^{-1}. \label{Ur_eqn}
\end{equation}
Substituting this expression for $U_r$ into Eq.~\eqref{f3f2eqn}, we have
\begin{equation}
\beta^2(F-1)^3 + 2\beta(F-1)^2 + (F-1) = 8\Wi^2r^2/L^2. \label{eq:Fminus1}
\end{equation}
Given that Eq.~(\ref{f3f2eqn}) implies $F-1\geq0$, a solution of Eq.~(\ref{eq:Fminus1}) reads 
\begin{equation}
F(r) = 1+(\zeta_1^{1/3} + \zeta_2^{1/3} -2)/(3\beta), \label{eq_F}
\end{equation}
where $\zeta_1 = a + \sqrt{a^2-1}$, $\zeta_2 = a - \sqrt{a^2 -1}$ and $a = 1 + 108\beta \Wi^2r^2/L^2$ with positive square-root and real cubic-root implied. Eventually, Eq.~\eqref{Ur_eqn} gives the velocity profile
\begin{equation}
U(r) = 2\int_0^1\frac{r'F(r')[1-\mathcal{U}(r-r')]}{\beta [F(r')-1]+1}\diff r', \label{U_eqn}
\end{equation}
where $\mathcal{U}(r)$ is the Heaviside step function. The set~\eqref{eq_C11}--\eqref{eq_C12}, along with Eqs.~\eqref{eq_F} \&~\eqref{U_eqn} give the complete FENE-P steady laminar pipe flow solution that we use to study the laminar drag. Their behaviors close to the pipe center, i.e., 
\begin{align}
U_r(r) &= -2r -16(1-\beta)\Wi^2L^{-2}r^3 + \cdots,\\ 
F(r) &= 1 + 8\Wi^2L^{-2}r^2 + \cdots,\\ 
C_{11}(r) &= 1 + 8\Wi^2(L^2-1)L^{-2}r^2 + \cdots, \\ 
C_{12}(r) &= -2\Wi r + 32\Wi^3L^{-2}r^3 + \cdots, \\
C_{22}(r) &= C_{33}(r) = 1 - 8\Wi^2L^{-2}r^2 + \cdots,
\end{align}
show that they exhibit even or odd symmetries with respect to $r$. These parities can assist when analyzing the symmetries of small perturbations close to the wall as performed for the case of Newtonian flows~\cite{priymak1998accurate} and when deploying them in the numerical calculations~\cite{meseguer2003linearized,malik2019linear}.

\begin{figure*}
\centering
\includegraphics[width=0.8\textwidth]{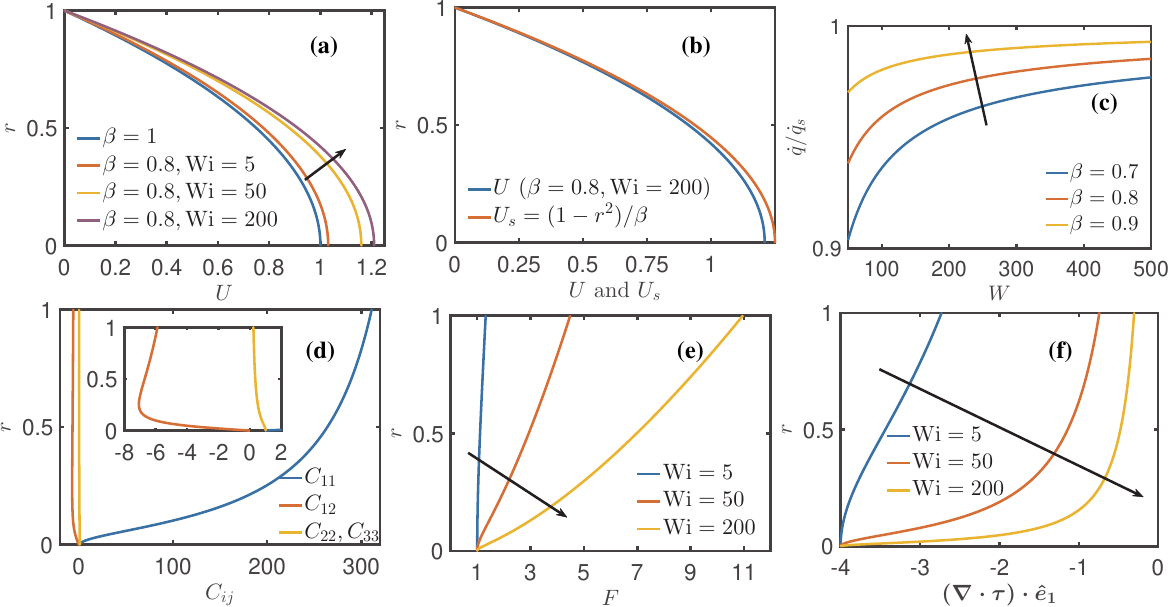}
	\caption{Profiles and flow rates for $L = 20$: (a) $U(r)$; 
	(b) $U(r)$ and $U_s(r)$; 
	(c) $\dot{q}/\dot{q}_s$ vs. $\Wi$; 
	(d) Components of $C_{ij}(r)$ for $\beta = 0.8$ and $\Wi = 50$ where the inset is 
	a zoom-in on $C_{12}$, and $C_{22}=C_{33}$; 
	(e) $F$ ($\beta = 0.8$); 
	(f) Divergence of polymer-stress for $\beta = 0.8$.
    The arrows, covering three profiles in each figure (a, c, e, f), indicate the increasing order of $\text{Wi}$ (a, e, f) or $\beta$ (c) on the curves.
	}
	\label{fig:mean_flow}
\end{figure*}

Figure~\ref{fig:mean_flow}(a) shows the velocity profiles in the Newtonian case ($\beta=1$) and three non-Newtonian cases ($\beta=0.8$ and increasing $\Wi$).
They reveal that under the same driving pressure gradient, the volume flow rate $\dot{q}=2\pi \int_0^1 U(r)r \diff r$  increases when $\beta$ drops from $1$ to $0.8$, thereby implying an apparent DR in the laminar regime. This increase in $\dot{q}$ is further amplified when considering increasingly large $\Wi$. However, this should not be interpreted as that the addition of polymer reduces drag compared to the pure solvent Newtonian case.

To better understand that, let us recall that the pressure gradient~\eqref{pressure-gradient} is defined based on the Reynolds number and therefore the total viscosity $\mu$. For a given fixed value of $\diff P/\diff x$, the total viscosity has to be the same for all values of $\beta$. Hence, when $\beta$ is altered, the viscosity of the solvent changes for the profiles shown in Fig.~\ref{fig:mean_flow}(a). This implies that the various profiles for different $\beta$ values correspond to different solvents, and that the non-Newtonian flow effectively exhibits less drag when compared against a Newtonian fluid---other than the solvent---having the same total viscosity as the one at the particular $\beta$ value considered.

We now show that the drag of a Newtonian flow of the pure solvent is lower than that of the non-Newtonian solution after adding polymers. In the non-Newtonian case, the dimensional pressure gradient is given by $\diff P^*/\diff x^* = -4\mu U_c/R^2$, where the superscript `$*$' indicates the dimensional value of a given variable. If the same dimensional pressure gradient is applied on the pure solvent, the centerline velocity increases by a factor of $1/\beta$ since $\diff P^*/\diff x^* =-4\mu_s [U_c/\beta]/R^2$. Hence the dimensional velocity profile of the pure solvent is $U_s^*=U_c(1-r^2)/\beta$ and its non-dimensional counterpart $U_s(r)$ is shown in Fig.~\ref{fig:mean_flow}(b) together with the profile  $U(r)$ of the polymeric solution. As can clearly be observed, the pure solvent experiences less drag. To study for a range of Weissenberg numbers $\Wi$, we compare the volume flow rate of the polymeric solution, $\dot{q} = 2\pi\int_0^1rU(r){\rm d}r$ against that of the pure solvent, $\dot{q}_s = \pi/(2\beta)$. As can be seen in Fig.~\ref{fig:mean_flow}(c), there is always a drag enhancement for all values of $\Wi$, and this result holds when varying $L$.

In the limit of $\Wi\rightarrow \infty$ or $L\rightarrow 0$, we find $F\rightarrow \infty$ from Eq.~(\ref{f3f2eqn}). Hence, considering the same limit in Eq.~\eqref{Ur_eqn}, we obtain $U_r\rightarrow -2r/\beta$, which is the derivative of $(1-r^2)/\beta$. These results therefore provide the bounds for the velocity profile as
\begin{equation}
1-r^2 \leq U(r) \leq (1-r^2)/\beta, \label{ineq_U}
\end{equation}
for all values of $\Wi$ and $L$. Physically, this implies that the polymeric solution experiences less drag than the Newtonian flow of a fluid having a viscosity identical to the total viscosity of the former, but higher drag than that of the pure solvent under same driving pressure gradient. This can be made more clearer from the behaviors of their respective friction factors. 

The Darcy friction factor is defined as 
\begin{equation}
f \equiv \frac{-8[\mu_sU_{r*}^* + (\mu-\mu_s)\tau_{12}^*]|_{r^*=R}}{\rho \langle U^* \rangle^2 }, \label{eq:f}
\end{equation}
where $\tau^*_{12}$ is a component of the dimensional version of the elastic stress tensor $\boldsymbol{\tau}$ and $\langle U^* \rangle$ is the cross-sectional average of $U^*(r^*)$. For Newtonian flows, the same definition holds but without the second term within the square brackets. The inequalities in Eq.~\eqref{ineq_U} translate into the following inequalities for the friction-factor:
\begin{equation}
\frac{64\beta^2}{\Re} 
\leq
\frac{4}{\Re}\left (\int_0^1U(r)r\diff r\right )^{-2}
\leq
\frac{64}{\Re}.
\label{ineq_f}
\end{equation}
The right-most term of Eq.~\eqref{ineq_f} is $f$ for a Newtonian fluid with viscosity same as the total viscosity of the solution. The middle and the left-most terms are $f$ for the polymer solution and the pure Newtonian solvent, respectively. These expressions for the friction-factor can be derived from Eqs.~\eqref{eq_C12},~\eqref{Ur_eqn} and~\eqref{eq:f} in the case of polymeric solutions and by using the parabolic profiles in the case of Newtonian flows. 

To analyze the influence of $\Wi$ and $L$, we consider the limit $\beta\rightarrow 0$ whereby the solutions are such that $F = 1+8\Wi^2r^2/L^2$, and $U(r) = 1-r^2 + 4\Wi^2(1-r^4)/L^2$. These relations show that the volume flow rate increases with $\Wi$ and decreases with $L$. Indeed, a large $\Wi$ implies that the relaxation time is far greater than the time scale of the flow, allowing the polymer strain---i.e. the stretching of the ends of the polymer molecules---to be increased by the mean shear. In what follows, we show that an enhanced stretching of polymers reduces the drag when compared with a Newtonian fluid with the same viscosity as the total viscosity of the non-Newtonian case. Moreover, a large $L$ implies that the ratio of $\sqrt{C_{ii}}$ to $L$ becomes small, resulting in a reduction of the restoring elastic modulus $F$. This, in turn, generates an increase in $C_{22}$ and $C_{33}$ given by Eq.~\eqref{eq_C22}. As shown below, an increase in $C_{22}$ yields an increase in the effective viscosity.

The profiles of the non-zero components of $C_{ij}(r)$ are shown in Fig.~\ref{fig:mean_flow}(d). The large values of $C_{11}$ originates from the polymers undergoing stretching in the axial direction. On the other hand, the fact that $C_{22}$ and $C_{33}$ have values below one implies contractions in both the radial and azimuthal directions. The term $C_{12}$ plays a crucial role in translating polymer strain represented by $C_{11}$ into an enhancement of the strain rate $U_r$ of the flow, as revealed by the $\boldsymbol{\hat{e}_1\hat{e}_1}$ component of Eq.~\eqref{C_eqn}: $2C_{12}U_r = (FC_{11}-1)/\Wi$. 

$F(r)$ is essentially a representation of the elastic modulus (see Fig.~\ref{fig:mean_flow}(e)). The polymers undergo maximum stretching near the wall given the higher values of the strain rate $U_r$. This causes an increase in $F$; a feature of FENE models  that is absent with the Oldroyd-B model. 

We can now explain the lower drag in the polymeric solution when compared to that of the Newtonian fluid with identical total viscosity. The divergence of the Newtonian stress tensor $(\boldsymbol{\nabla}\cdot\boldsymbol{\tau})\cdot\boldsymbol{\hat{e}_1}$ (see Fig.~\ref{fig:mean_flow}(f)) without polymer stands at a constant value of $-4$, with the negative sign implying (positive) dissipation. When polymers are added, a ratio of $(1-\beta)$ of this viscous dissipative component is replaced by the component due to polymer stretching, which is also dissipative of the fluid momentum. As evident from Fig.~\ref{fig:mean_flow}(f), this component is always greater than $-4$ throughout the flow field except at the center of the pipe. (Note that this component can be written as $\{-2r^2/[\beta(F-1)+1]\}_r/r$ which takes the value of $-4$ in the limit $r\rightarrow 0$, which is same as the Newtonian component.) To counterbalance the constant pressure gradient term, the Newtonian part readjusts itself to increase the dissipation by increasing the overall strain rate, thereby resulting in an enhanced volume flow rate. Such an increase in the strain rate is characterized by an increase in the slope of the velocity near wall with respect to $1-r$ in Fig.~\ref{fig:mean_flow}(a). 

To explain the decrease in $|(\boldsymbol{\nabla}\cdot\boldsymbol{\tau})\cdot\boldsymbol{\hat{e}_1}|$ as $r$ increases, we first note that $C_{12} \equiv \langle\tilde{R}_1\tilde{R}_2 \rangle = 0$ at the pipe center, thus implying that the polymer molecules undergo an uncorrelated random motion there. Since $C_{11}$ increases from the value of $1.0$ with respect to $r$, the polymers undergo a rapid stretching in the $x$-direction (see Fig.~\ref{fig:mean_flow}(d)) causing the $\tilde{R}_2$ component to pick-up a negative correlation with the $\tilde{R}_1$ component that undergo stretching. This negative correlation is due to the resistance to elongation associated with the restoring tendency $F$. However, this negative correlation starts to decrease with respect to $r$ partly due to the slow-down in the axial stretching $C_{11}$ and the restoring tendency, $F$---i.e. $C_{11rr}<0$ and $F_{rr}<0$ for most part of the pipe. This causes less drag as the absolute value of the resistance term $[(C_{12}F)_r+C_{12}F/r]/\Wi$ comes down. 

This phenomenon of $C_{11}$ or $C_{12}$ playing a crucial role in the laminar regime is in stark contrast with the turbulent regime, where the transverse diagonal component $C_{22}$ (or $C_{33}$) plays a critical role---it acquires a radial distribution similar to the effective viscosity~\cite{benzi2010short, benzi2006maximum}. In fact, both interpretations are reconciled as explained below.

\begin{figure}\centering
\includegraphics[width=0.35\textwidth]{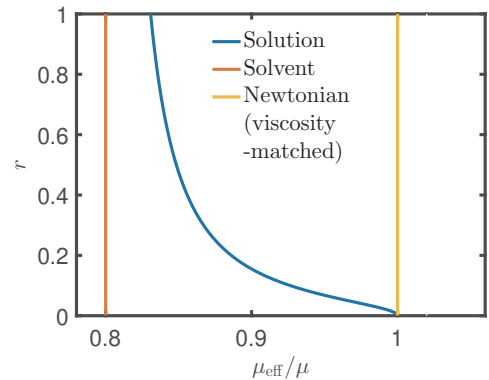}
	\caption{Effective viscosity for $L = 20$, $\beta = 0.8$ and $\Wi = 50$}
	\label{fig:nu_eff}
\end{figure}

The decrease in flow rate when compared to that of the pure solvent, or, the increase of the same when compared to that of a viscosity-matched Newtonian fluid, can also be understood by considering the effective viscosity $\mu_\text{eff}$, which is defined as 
\begin{align}
\frac{\mu_\text{eff}(r)}{\mu} &= \frac{[{\beta}\nabla^2\boldsymbol{u} + {(1-\beta)}\boldsymbol{\nabla}\cdot\boldsymbol{\tau}]\cdot\boldsymbol{\hat{e}_1}}{[\nabla^2\boldsymbol{u}]\cdot\boldsymbol{\hat{e}_1}} \nonumber \\
&= \beta + (1-\beta)\left [ \frac{1}{F} - \frac{rU_rF_r}{F^2(U_r+rU_{rr})} \right]. \label{eq_nu}
\end{align}
Taking note of the facts that $F\rightarrow \infty$ in the limits $\Wi\rightarrow \infty$ or $L\rightarrow 0$, and  $F\rightarrow 1$ in the limits $\Wi\rightarrow 0$ or $L\rightarrow \infty$, we arrive at the following bounds for $\mu_\text{eff}/\mu$:
\begin{equation}
\beta \leq \mu_\text{eff}/\mu \leq 1, \label{ineq_mu}
\end{equation}
which contains the analogous information as the inequalities in Eq.~\eqref{ineq_U}, i.e., the effective viscosity is less than the total viscosity, but higher than the viscosity of the solvent. In Fig.~\ref{fig:nu_eff}, $\mu_\text{eff}/\mu$ is shown for parameters set equal to that of Fig.~\ref{fig:mean_flow}(d).  
The figure confirms that the ratio $\mu_\text{eff}(r)/\mu$ obeys the inequalities in Eq.~\eqref{ineq_mu}. Clearly, $\mu_\text{eff}(r)/\mu$ has a trend with respect to $r$ that is opposite to that of $C_{11}$. This shows that the stretching in the axial direction reduces the effective viscosity. As a general non-auxetic matter, the polymer exhibits opposite trend in $C_{22}$ and $C_{33}$ by way of contraction of the polymer dumbbells. This sets $C_{22}$ and $C_{33}$ to follow the same trend as $\mu_\text{eff}(r)/\mu$ (see inset of Fig.~\ref{fig:mean_flow}(d)). As stated previously, this phenomenon is well-known in the turbulent regime~\cite{benzi2010short,benzi2006maximum}. The linear relation (up to the leading order) between of $\mu_\text{eff}(r)/\mu$ and $C_{22}$ is revealed by re-writing Eq.~(\ref{eq_nu}) as
\begin{equation}
\frac{\mu_\text{eff}(r)}{\mu} = \beta + (1-\beta)C_{22} \left \{1 - \frac{2rF_r}{2F-U_{rr}[\beta(F-1)+1]}\right \}, \nonumber
\end{equation} 
where the second term within the braces can be shown to be of $O(r^2)$.

In summary, the laminar regime of this non-Newtonian flow has interesting bounds for flow and friction determined by two corresponding Newtonian flows. The flow of a polymeric solution with a total viscosity identical to that of a Newtonian fluid exhibits lower drag on comparison with the latter. Nonetheless, the experienced drag is higher than that of the pure solvent under the same driving pressure gradient. It should be noted that these bounds are also valid for plane Poiseuille flow. 

During the course arriving these bounds, we also solved the FENE-P model in a form widely known to the community researching turbulence and transition. These alternative solutions to those found by Cruz \emph{et al.}~\cite{cruz2005analytical} will enable the study of small perturbations to FENE-P model expressed in Eqs.~(\ref{gov_eqn_u}) and~(\ref{gov_eqn_c}). However, any future study on transition should take into account that the stability of the viscoelastic flow at a certain $\Re$ should be compared against that of the corresponding flow of the solvent at $\Re/\beta$ in order to maintain the same pressure gradient. 
If not, the comparison would not be against stability of the flow of the solvent, but with another Newtonian fluid with the same viscosity as the total viscosity of the solution.

This rich phenomenology can also be interpreted as an explanation of the well-known shear-thinning effect through the following sequence of comprehensive arguments. (1) The strain rate of the flow stretches the polymers in the axial direction, a phenomenon also known from earlier literature; (2) Such an extension causes contraction in the radial direction, due to the resistance to elongation exerted by restoring force between the ends of the polymer molecules; (3) This triggers a nonzero negative correlation between the $x-$component and $r-$component of the end-to-end polymer vector, resulting in an elastic shear stress ($\tau_{12}$) to develop; (4) Since there is no strain rate at the pipe center, this correlation is zero due to the liberty of the ends of the polymers to be at random motion; (5) Due the increase of the elastic modulus with respect to $r$, a feature of the FENE-P model, the stretching rate with respect to $r$ slows down, giving rise to convexity of $C_{11}$ (i.e., $C_{11rr}<0$), which in turn causes concavity in $\tau_{12}$; (6) The points in (3), (4) and (5) imply that the divergence of elastic shear-stress, i.e, the resistive force contribution to the flow by the polymers decreases with $r$; (7) The driving force, i.e., the pressure gradient, which is balanced completely by the Newtonian viscous force and elastic force, is a constant; (8) Since the elastic force decrease radially, the Newtonian viscous force should increase with $r$;  (9) Finally, this is achieved by enhancing strain rate owing to the fact that the dynamic viscosity of the solvent is a constant, thus increasing the flow rate. The last point in this sequence could be perceived as shear-thinning by introducing a variable viscosity as a factoring function together with $\nabla^2 U$, which balances the total dissipation.

Indeed, we proved that $U$ and the components of the tensor $\textbf{\textit{\textsf{C}}}$ exhibit even or odd symmetries with respect to $r$. As an outlook, these symmetries can be exploited to derive conditions for the regularity of perturbations at the pipe center as derived in the Newtonian case~\cite{malik2019linear}. These are expected to be valuable for the study of the stability and transition of viscoelastic pipe flows. 

We thank Dr. Jurriaan J. J. Gillissen at the Department of Mathematics, University College London, for his fruitful comments on this letter.
\bibliography{mbs_final}

\begin{thebibliography}{32}%
\makeatletter
\providecommand \@ifxundefined [1]{%
 \@ifx{#1\undefined}
}%
\providecommand \@ifnum [1]{%
 \ifnum #1\expandafter \@firstoftwo
 \else \expandafter \@secondoftwo
 \fi
}%
\providecommand \@ifx [1]{%
 \ifx #1\expandafter \@firstoftwo
 \else \expandafter \@secondoftwo
 \fi
}%
\providecommand \natexlab [1]{#1}%
\providecommand \enquote  [1]{``#1''}%
\providecommand \bibnamefont  [1]{#1}%
\providecommand \bibfnamefont [1]{#1}%
\providecommand \citenamefont [1]{#1}%
\providecommand \href@noop [0]{\@secondoftwo}%
\providecommand \href [0]{\begingroup \@sanitize@url \@href}%
\providecommand \@href[1]{\@@startlink{#1}\@@href}%
\providecommand \@@href[1]{\endgroup#1\@@endlink}%
\providecommand \@sanitize@url [0]{\catcode `\\12\catcode `\$12\catcode
  `\&12\catcode `\#12\catcode `\^12\catcode `\_12\catcode `\%12\relax}%
\providecommand \@@startlink[1]{}%
\providecommand \@@endlink[0]{}%
\providecommand \url  [0]{\begingroup\@sanitize@url \@url }%
\providecommand \@url [1]{\endgroup\@href {#1}{\urlprefix }}%
\providecommand \urlprefix  [0]{URL }%
\providecommand \Eprint [0]{\href }%
\providecommand \doibase [0]{https://doi.org/}%
\providecommand \selectlanguage [0]{\@gobble}%
\providecommand \bibinfo  [0]{\@secondoftwo}%
\providecommand \bibfield  [0]{\@secondoftwo}%
\providecommand \translation [1]{[#1]}%
\providecommand \BibitemOpen [0]{}%
\providecommand \bibitemStop [0]{}%
\providecommand \bibitemNoStop [0]{.\EOS\space}%
\providecommand \EOS [0]{\spacefactor3000\relax}%
\providecommand \BibitemShut  [1]{\csname bibitem#1\endcsname}%
\let\auto@bib@innerbib\@empty
\bibitem [{\citenamefont {Dixon}, \citenamefont {Lucey},\ and\ \citenamefont
  {Carpenter}(1994)}]{dixon1994optimization}%
  \BibitemOpen
  \bibfield  {author} {\bibinfo {author} {\bibfnamefont {A.~E.}\ \bibnamefont
  {Dixon}}, \bibinfo {author} {\bibfnamefont {A.~D.}\ \bibnamefont {Lucey}},\
  and\ \bibinfo {author} {\bibfnamefont {P.~W.}\ \bibnamefont {Carpenter}},\
  }\bibfield  {title} {\enquote {\bibinfo {title} {Optimization of viscoelastic
  compliant walls for transition delay},}\ }\href@noop {} {\bibfield  {journal}
  {\bibinfo  {journal} {AIAA J.}\ }\textbf {\bibinfo {volume} {32}},\ \bibinfo
  {pages} {256--267} (\bibinfo {year} {1994})}\BibitemShut {NoStop}%
\bibitem [{\citenamefont {Carpenter}, \citenamefont {Lucey},\ and\
  \citenamefont {Davies}(2001)}]{carpenter2001progress}%
  \BibitemOpen
  \bibfield  {author} {\bibinfo {author} {\bibfnamefont {P.~W.}\ \bibnamefont
  {Carpenter}}, \bibinfo {author} {\bibfnamefont {A.~D.}\ \bibnamefont
  {Lucey}},\ and\ \bibinfo {author} {\bibfnamefont {C.}~\bibnamefont
  {Davies}},\ }\bibfield  {title} {\enquote {\bibinfo {title} {Progress on the
  use of compliant walls for laminar-flow control},}\ }\href@noop {} {\bibfield
   {journal} {\bibinfo  {journal} {J. Aircraft}\ }\textbf {\bibinfo {volume}
  {38}},\ \bibinfo {pages} {504--512} (\bibinfo {year} {2001})}\BibitemShut
  {NoStop}%
\bibitem [{\citenamefont {Malik}, \citenamefont {Skote},\ and\ \citenamefont
  {Bouffanais}(2018)}]{malik2018growth}%
  \BibitemOpen
  \bibfield  {author} {\bibinfo {author} {\bibfnamefont {M.}~\bibnamefont
  {Malik}}, \bibinfo {author} {\bibfnamefont {M.}~\bibnamefont {Skote}},\ and\
  \bibinfo {author} {\bibfnamefont {R.}~\bibnamefont {Bouffanais}},\ }\bibfield
   {title} {\enquote {\bibinfo {title} {Growth mechanisms of perturbations in
  boundary layers over a compliant wall},}\ }\href@noop {} {\bibfield
  {journal} {\bibinfo  {journal} {Phys. Rev. Fluids}\ }\textbf {\bibinfo
  {volume} {3}},\ \bibinfo {pages} {013903} (\bibinfo {year}
  {2018})}\BibitemShut {NoStop}%
\bibitem [{\citenamefont {Toms}(1948)}]{toms1948some}%
  \BibitemOpen
  \bibfield  {author} {\bibinfo {author} {\bibfnamefont {B.~A.}\ \bibnamefont
  {Toms}},\ }\bibfield  {title} {\enquote {\bibinfo {title} {Some observations
  on the flow of linear polymer solutions through straight tubes at large
  reynolds numbers},}\ }\href@noop {} {\bibfield  {journal} {\bibinfo
  {journal} {Proc. of 1st Int. Cong. On Rheology, 1948}\ }\textbf {\bibinfo
  {volume} {135}} (\bibinfo {year} {1948})}\BibitemShut {NoStop}%
\bibitem [{\citenamefont {Sureshkumar}\ and\ \citenamefont
  {Beris}(1995)}]{sureshkumar1995linear}%
  \BibitemOpen
  \bibfield  {author} {\bibinfo {author} {\bibfnamefont {R.}~\bibnamefont
  {Sureshkumar}}\ and\ \bibinfo {author} {\bibfnamefont {A.~N.}\ \bibnamefont
  {Beris}},\ }\bibfield  {title} {\enquote {\bibinfo {title} {Linear stability
  analysis of viscoelastic poiseuille flow using an arnoldi-based
  orthogonalization algorithm},}\ }\href@noop {} {\bibfield  {journal}
  {\bibinfo  {journal} {J. Non-Newtonian Fluid Mech.}\ }\textbf {\bibinfo
  {volume} {56}},\ \bibinfo {pages} {151--182} (\bibinfo {year}
  {1995})}\BibitemShut {NoStop}%
\bibitem [{\citenamefont {Zhang}\ \emph {et~al.}(2013)\citenamefont {Zhang},
  \citenamefont {Lashgari}, \citenamefont {Zaki},\ and\ \citenamefont
  {Brandt}}]{zhang2013linear}%
  \BibitemOpen
  \bibfield  {author} {\bibinfo {author} {\bibfnamefont {M.}~\bibnamefont
  {Zhang}}, \bibinfo {author} {\bibfnamefont {I.}~\bibnamefont {Lashgari}},
  \bibinfo {author} {\bibfnamefont {T.~A.}\ \bibnamefont {Zaki}},\ and\
  \bibinfo {author} {\bibfnamefont {L.}~\bibnamefont {Brandt}},\ }\bibfield
  {title} {\enquote {\bibinfo {title} {Linear stability analysis of channel
  flow of viscoelastic {O}ldroyd-{B} and {FENE-P} fluids},}\ }\href@noop {}
  {\bibfield  {journal} {\bibinfo  {journal} {J. Fluid Mech.}\ }\textbf
  {\bibinfo {volume} {737}},\ \bibinfo {pages} {249--279} (\bibinfo {year}
  {2013})}\BibitemShut {NoStop}%
\bibitem [{\citenamefont {Garg}\ \emph {et~al.}(2018)\citenamefont {Garg},
  \citenamefont {Chaudhary}, \citenamefont {Khalid}, \citenamefont {Shankar},\
  and\ \citenamefont {Subramanian}}]{garg2018viscoelastic}%
  \BibitemOpen
  \bibfield  {author} {\bibinfo {author} {\bibfnamefont {P.}~\bibnamefont
  {Garg}}, \bibinfo {author} {\bibfnamefont {I.}~\bibnamefont {Chaudhary}},
  \bibinfo {author} {\bibfnamefont {M.}~\bibnamefont {Khalid}}, \bibinfo
  {author} {\bibfnamefont {V.}~\bibnamefont {Shankar}},\ and\ \bibinfo {author}
  {\bibfnamefont {G.}~\bibnamefont {Subramanian}},\ }\bibfield  {title}
  {\enquote {\bibinfo {title} {Viscoelastic pipe flow is linearly unstable},}\
  }\href@noop {} {\bibfield  {journal} {\bibinfo  {journal} {Phys. Rev. Lett.}\
  }\textbf {\bibinfo {volume} {121}},\ \bibinfo {pages} {024502} (\bibinfo
  {year} {2018})}\BibitemShut {NoStop}%
\bibitem [{\citenamefont {Brandi}, \citenamefont {Mendon{\c{c}}a},\ and\
  \citenamefont {Souza}(2019)}]{brandi2019dns}%
  \BibitemOpen
  \bibfield  {author} {\bibinfo {author} {\bibfnamefont {A.}~\bibnamefont
  {Brandi}}, \bibinfo {author} {\bibfnamefont {M.}~\bibnamefont
  {Mendon{\c{c}}a}},\ and\ \bibinfo {author} {\bibfnamefont {L.}~\bibnamefont
  {Souza}},\ }\bibfield  {title} {\enquote {\bibinfo {title} {{DNS} and {LST}
  stability analysis of {O}ldroyd-{B} fluid in a flow between two parallel
  plates},}\ }\href@noop {} {\bibfield  {journal} {\bibinfo  {journal} {J.
  Non-Newtonian Fluid Mech.}\ }\textbf {\bibinfo {volume} {267}},\ \bibinfo
  {pages} {14--27} (\bibinfo {year} {2019})}\BibitemShut {NoStop}%
\bibitem [{\citenamefont {White}\ and\ \citenamefont
  {Mungal}(2008)}]{white2008mechanics}%
  \BibitemOpen
  \bibfield  {author} {\bibinfo {author} {\bibfnamefont {C.~M.}\ \bibnamefont
  {White}}\ and\ \bibinfo {author} {\bibfnamefont {M.~G.}\ \bibnamefont
  {Mungal}},\ }\bibfield  {title} {\enquote {\bibinfo {title} {Mechanics and
  prediction of turbulent drag reduction with polymer additives},}\ }\href@noop
  {} {\bibfield  {journal} {\bibinfo  {journal} {Annu. Rev. Fluid Mech.}\
  }\textbf {\bibinfo {volume} {40}},\ \bibinfo {pages} {235--256} (\bibinfo
  {year} {2008})}\BibitemShut {NoStop}%
\bibitem [{\citenamefont {Graham}(2004)}]{graham2004drag}%
  \BibitemOpen
  \bibfield  {author} {\bibinfo {author} {\bibfnamefont {M.~D.}\ \bibnamefont
  {Graham}},\ }\bibfield  {title} {\enquote {\bibinfo {title} {Drag reduction
  in turbulent flow of polymer solutions},}\ }\href@noop {} {\bibfield
  {journal} {\bibinfo  {journal} {Rheology Rev.}\ }\textbf {\bibinfo {volume}
  {2}},\ \bibinfo {pages} {143--170} (\bibinfo {year} {2004})}\BibitemShut
  {NoStop}%
\bibitem [{\citenamefont {Samanta}\ \emph {et~al.}(2013)\citenamefont
  {Samanta}, \citenamefont {Dubief}, \citenamefont {Holzner}, \citenamefont
  {Sch{\"a}fer}, \citenamefont {Morozov}, \citenamefont {Wagner},\ and\
  \citenamefont {Hof}}]{samanta2013elasto}%
  \BibitemOpen
  \bibfield  {author} {\bibinfo {author} {\bibfnamefont {D.}~\bibnamefont
  {Samanta}}, \bibinfo {author} {\bibfnamefont {Y.}~\bibnamefont {Dubief}},
  \bibinfo {author} {\bibfnamefont {M.}~\bibnamefont {Holzner}}, \bibinfo
  {author} {\bibfnamefont {C.}~\bibnamefont {Sch{\"a}fer}}, \bibinfo {author}
  {\bibfnamefont {A.~N.}\ \bibnamefont {Morozov}}, \bibinfo {author}
  {\bibfnamefont {C.}~\bibnamefont {Wagner}},\ and\ \bibinfo {author}
  {\bibfnamefont {B.}~\bibnamefont {Hof}},\ }\bibfield  {title} {\enquote
  {\bibinfo {title} {Elasto-inertial turbulence},}\ }\href@noop {} {\bibfield
  {journal} {\bibinfo  {journal} {Proceedings of the National Academy of
  Sciences}\ }\textbf {\bibinfo {volume} {110}},\ \bibinfo {pages}
  {10557--10562} (\bibinfo {year} {2013})}\BibitemShut {NoStop}%
\bibitem [{\citenamefont {Dubief}, \citenamefont {Terrapon},\ and\
  \citenamefont {Soria}(2013)}]{dubief2013mechanism}%
  \BibitemOpen
  \bibfield  {author} {\bibinfo {author} {\bibfnamefont {Y.}~\bibnamefont
  {Dubief}}, \bibinfo {author} {\bibfnamefont {V.~E.}\ \bibnamefont
  {Terrapon}},\ and\ \bibinfo {author} {\bibfnamefont {J.}~\bibnamefont
  {Soria}},\ }\bibfield  {title} {\enquote {\bibinfo {title} {On the mechanism
  of elasto-inertial turbulence},}\ }\href@noop {} {\bibfield  {journal}
  {\bibinfo  {journal} {Physics of Fluids}\ }\textbf {\bibinfo {volume} {25}},\
  \bibinfo {pages} {110817} (\bibinfo {year} {2013})}\BibitemShut {NoStop}%
\bibitem [{\citenamefont {Lumley}(1969)}]{lumley1969drag}%
  \BibitemOpen
  \bibfield  {author} {\bibinfo {author} {\bibfnamefont {J.~L.}\ \bibnamefont
  {Lumley}},\ }\bibfield  {title} {\enquote {\bibinfo {title} {Drag reduction
  by additives},}\ }\href@noop {} {\bibfield  {journal} {\bibinfo  {journal}
  {Ann. Rev. Fluid Mech.}\ }\textbf {\bibinfo {volume} {1}},\ \bibinfo {pages}
  {367--384} (\bibinfo {year} {1969})}\BibitemShut {NoStop}%
\bibitem [{\citenamefont {Nieuwstadt}\ and\ \citenamefont
  {Den~Toonder}(2001)}]{nieuwstadt2001drag}%
  \BibitemOpen
  \bibfield  {author} {\bibinfo {author} {\bibfnamefont {F.}~\bibnamefont
  {Nieuwstadt}}\ and\ \bibinfo {author} {\bibfnamefont {J.}~\bibnamefont
  {Den~Toonder}},\ }\bibfield  {title} {\enquote {\bibinfo {title} {Drag
  reduction by additives: a review},}\ }in\ \href@noop {} {\emph {\bibinfo
  {booktitle} {Turbulence Structure and Modulation}}}\ (\bibinfo  {publisher}
  {Springer},\ \bibinfo {year} {2001})\ pp.\ \bibinfo {pages}
  {269--316}\BibitemShut {NoStop}%
\bibitem [{\citenamefont {Sreenivasan}\ and\ \citenamefont
  {White}(2000)}]{sreenivasan2000onset}%
  \BibitemOpen
  \bibfield  {author} {\bibinfo {author} {\bibfnamefont {K.~R.}\ \bibnamefont
  {Sreenivasan}}\ and\ \bibinfo {author} {\bibfnamefont {C.~M.}\ \bibnamefont
  {White}},\ }\bibfield  {title} {\enquote {\bibinfo {title} {The onset of drag
  reduction by dilute polymer additives, and the maximum drag reduction
  asymptote},}\ }\href@noop {} {\bibfield  {journal} {\bibinfo  {journal} {J.
  Fluid Mech.}\ }\textbf {\bibinfo {volume} {409}},\ \bibinfo {pages}
  {149--164} (\bibinfo {year} {2000})}\BibitemShut {NoStop}%
\bibitem [{\citenamefont {Virk}\ \emph {et~al.}(1967)\citenamefont {Virk},
  \citenamefont {Merrill}, \citenamefont {Mickley}, \citenamefont {Smith},\
  and\ \citenamefont {Mollo-Christensen}}]{virk1967toms}%
  \BibitemOpen
  \bibfield  {author} {\bibinfo {author} {\bibfnamefont {P.~S.}\ \bibnamefont
  {Virk}}, \bibinfo {author} {\bibfnamefont {E.}~\bibnamefont {Merrill}},
  \bibinfo {author} {\bibfnamefont {H.}~\bibnamefont {Mickley}}, \bibinfo
  {author} {\bibfnamefont {K.}~\bibnamefont {Smith}},\ and\ \bibinfo {author}
  {\bibfnamefont {E.}~\bibnamefont {Mollo-Christensen}},\ }\bibfield  {title}
  {\enquote {\bibinfo {title} {The toms phenomenon: turbulent pipe flow of
  dilute polymer solutions},}\ }\href@noop {} {\bibfield  {journal} {\bibinfo
  {journal} {J. Fluid Mech.}\ }\textbf {\bibinfo {volume} {30}},\ \bibinfo
  {pages} {305--328} (\bibinfo {year} {1967})}\BibitemShut {NoStop}%
\bibitem [{\citenamefont {Kenis}\ and\ \citenamefont
  {Hoyt}(1971)}]{kenis1971friction}%
  \BibitemOpen
  \bibfield  {author} {\bibinfo {author} {\bibfnamefont {P.~R.}\ \bibnamefont
  {Kenis}}\ and\ \bibinfo {author} {\bibfnamefont {J.}~\bibnamefont {Hoyt}},\
  }\href@noop {} {\enquote {\bibinfo {title} {Friction reduction by algal and
  bacterial polymers},}\ }\bibinfo {type} {Tech. Rep.}\ (\bibinfo
  {institution} {Naval Undersea Research And Development Centre, San Diego,
  CA},\ \bibinfo {year} {1971})\BibitemShut {NoStop}%
\bibitem [{\citenamefont {Benzi}\ \emph {et~al.}(2006)\citenamefont {Benzi},
  \citenamefont {De~Angelis}, \citenamefont {L'vov}, \citenamefont
  {Procaccia},\ and\ \citenamefont {Tiberkevich}}]{benzi2006maximum}%
  \BibitemOpen
  \bibfield  {author} {\bibinfo {author} {\bibfnamefont {R.}~\bibnamefont
  {Benzi}}, \bibinfo {author} {\bibfnamefont {E.}~\bibnamefont {De~Angelis}},
  \bibinfo {author} {\bibfnamefont {V.}~\bibnamefont {L'vov}}, \bibinfo
  {author} {\bibfnamefont {I.}~\bibnamefont {Procaccia}},\ and\ \bibinfo
  {author} {\bibfnamefont {V.}~\bibnamefont {Tiberkevich}},\ }\bibfield
  {title} {\enquote {\bibinfo {title} {Maximum drag reduction asymptotes and
  the cross-over to the newtonian plug},}\ }\href@noop {} {\bibfield  {journal}
  {\bibinfo  {journal} {J. Fluid Mech.}\ }\textbf {\bibinfo {volume} {551}},\
  \bibinfo {pages} {185--195} (\bibinfo {year} {2006})}\BibitemShut {NoStop}%
\bibitem [{\citenamefont {Xi}\ and\ \citenamefont
  {Graham}(2012)}]{xi2012dynamics}%
  \BibitemOpen
  \bibfield  {author} {\bibinfo {author} {\bibfnamefont {L.}~\bibnamefont
  {Xi}}\ and\ \bibinfo {author} {\bibfnamefont {M.~D.}\ \bibnamefont
  {Graham}},\ }\bibfield  {title} {\enquote {\bibinfo {title} {Dynamics on the
  laminar-turbulent boundary and the origin of the maximum drag reduction
  asymptote},}\ }\href@noop {} {\bibfield  {journal} {\bibinfo  {journal}
  {Phys. Rev. Lett.}\ }\textbf {\bibinfo {volume} {108}},\ \bibinfo {pages}
  {028301} (\bibinfo {year} {2012})}\BibitemShut {NoStop}%
\bibitem [{\citenamefont {Lopez}, \citenamefont {Choueiri},\ and\ \citenamefont
  {Hof}(2019)}]{lopez2019dynamics}%
  \BibitemOpen
  \bibfield  {author} {\bibinfo {author} {\bibfnamefont {J.~M.}\ \bibnamefont
  {Lopez}}, \bibinfo {author} {\bibfnamefont {G.~H.}\ \bibnamefont
  {Choueiri}},\ and\ \bibinfo {author} {\bibfnamefont {B.}~\bibnamefont
  {Hof}},\ }\bibfield  {title} {\enquote {\bibinfo {title} {Dynamics of
  viscoelastic pipe flow at low reynolds numbers in the maximum drag reduction
  limit},}\ }\href@noop {} {\bibfield  {journal} {\bibinfo  {journal} {J. Fluid
  Mech.}\ }\textbf {\bibinfo {volume} {874}},\ \bibinfo {pages} {699--719}
  (\bibinfo {year} {2019})}\BibitemShut {NoStop}%
\bibitem [{\citenamefont {Choueiri}, \citenamefont {Lopez},\ and\ \citenamefont
  {Hof}(2018)}]{choueiri2018exceeding}%
  \BibitemOpen
  \bibfield  {author} {\bibinfo {author} {\bibfnamefont {G.~H.}\ \bibnamefont
  {Choueiri}}, \bibinfo {author} {\bibfnamefont {J.~M.}\ \bibnamefont
  {Lopez}},\ and\ \bibinfo {author} {\bibfnamefont {B.}~\bibnamefont {Hof}},\
  }\bibfield  {title} {\enquote {\bibinfo {title} {Exceeding the asymptotic
  limit of polymer drag reduction},}\ }\href@noop {} {\bibfield  {journal}
  {\bibinfo  {journal} {Phys. Rev. Lett.}\ }\textbf {\bibinfo {volume} {120}},\
  \bibinfo {pages} {124501} (\bibinfo {year} {2018})}\BibitemShut {NoStop}%
\bibitem [{\citenamefont {Pereira}, \citenamefont {Thompson},\ and\
  \citenamefont {Mompean}(2019)}]{pereira2019beyond}%
  \BibitemOpen
  \bibfield  {author} {\bibinfo {author} {\bibfnamefont {A.}~\bibnamefont
  {Pereira}}, \bibinfo {author} {\bibfnamefont {R.~L.}\ \bibnamefont
  {Thompson}},\ and\ \bibinfo {author} {\bibfnamefont {G.}~\bibnamefont
  {Mompean}},\ }\bibfield  {title} {\enquote {\bibinfo {title} {Beyond the
  maximum drag reduction asymptote: the pseudo-laminar state},}\ }\href@noop {}
  {\bibfield  {journal} {\bibinfo  {journal} {arXiv preprint arXiv:1911.00439}\
  } (\bibinfo {year} {2019})}\BibitemShut {NoStop}%
\bibitem [{\citenamefont {Perlekar}, \citenamefont {Mitra},\ and\ \citenamefont
  {Pandit}(2006)}]{perlekar2006manifestations}%
  \BibitemOpen
  \bibfield  {author} {\bibinfo {author} {\bibfnamefont {P.}~\bibnamefont
  {Perlekar}}, \bibinfo {author} {\bibfnamefont {D.}~\bibnamefont {Mitra}},\
  and\ \bibinfo {author} {\bibfnamefont {R.}~\bibnamefont {Pandit}},\
  }\bibfield  {title} {\enquote {\bibinfo {title} {Manifestations of drag
  reduction by polymer additives in decaying, homogeneous, isotropic
  turbulence},}\ }\href@noop {} {\bibfield  {journal} {\bibinfo  {journal}
  {Phys. Rev. Lett.}\ }\textbf {\bibinfo {volume} {97}},\ \bibinfo {pages}
  {264501} (\bibinfo {year} {2006})}\BibitemShut {NoStop}%
\bibitem [{\citenamefont {Benzi}(2010)}]{benzi2010short}%
  \BibitemOpen
  \bibfield  {author} {\bibinfo {author} {\bibfnamefont {R.}~\bibnamefont
  {Benzi}},\ }\bibfield  {title} {\enquote {\bibinfo {title} {A short review on
  drag reduction by polymers in wall bounded turbulence},}\ }\href@noop {}
  {\bibfield  {journal} {\bibinfo  {journal} {Physica D: Nonlinear Phenomena}\
  }\textbf {\bibinfo {volume} {239}},\ \bibinfo {pages} {1338--1345} (\bibinfo
  {year} {2010})}\BibitemShut {NoStop}%
\bibitem [{\citenamefont {Phan-Thien}\ and\ \citenamefont
  {Mai-Duy}(2017)}]{phan2017understanding}%
  \BibitemOpen
  \bibfield  {author} {\bibinfo {author} {\bibfnamefont {N.}~\bibnamefont
  {Phan-Thien}}\ and\ \bibinfo {author} {\bibfnamefont {N.}~\bibnamefont
  {Mai-Duy}},\ }\href@noop {} {\emph {\bibinfo {title} {Understanding
  Viscoelasticity}}}\ (\bibinfo  {publisher} {Springer},\ \bibinfo {address}
  {Berlin},\ \bibinfo {year} {2017})\BibitemShut {NoStop}%
\bibitem [{\citenamefont {Bird}, \citenamefont {Stewart},\ and\ \citenamefont
  {Lightfoot}(2007)}]{bird2007transport}%
  \BibitemOpen
  \bibfield  {author} {\bibinfo {author} {\bibfnamefont {R.~B.}\ \bibnamefont
  {Bird}}, \bibinfo {author} {\bibfnamefont {W.~E.}\ \bibnamefont {Stewart}},\
  and\ \bibinfo {author} {\bibfnamefont {E.~N.}\ \bibnamefont {Lightfoot}},\
  }\href@noop {} {\emph {\bibinfo {title} {Transport phenomena}}}\ (\bibinfo
  {publisher} {John Wiley},\ \bibinfo {year} {2007})\BibitemShut {NoStop}%
\bibitem [{\citenamefont {Cruz}, \citenamefont {Pinho},\ and\ \citenamefont
  {Oliveira}(2005)}]{cruz2005analytical}%
  \BibitemOpen
  \bibfield  {author} {\bibinfo {author} {\bibfnamefont {D.}~\bibnamefont
  {Cruz}}, \bibinfo {author} {\bibfnamefont {F.}~\bibnamefont {Pinho}},\ and\
  \bibinfo {author} {\bibfnamefont {P.~J.}\ \bibnamefont {Oliveira}},\
  }\bibfield  {title} {\enquote {\bibinfo {title} {Analytical solutions for
  fully developed laminar flow of some viscoelastic liquids with a newtonian
  solvent contribution},}\ }\href@noop {} {\bibfield  {journal} {\bibinfo
  {journal} {Journal of Non-Newtonian Fluid Mechanics}\ }\textbf {\bibinfo
  {volume} {132}},\ \bibinfo {pages} {28--35} (\bibinfo {year}
  {2005})}\BibitemShut {NoStop}%
\bibitem [{\citenamefont {Oliveira}(2002)}]{oliveira2002exact}%
  \BibitemOpen
  \bibfield  {author} {\bibinfo {author} {\bibfnamefont {P.~J.}\ \bibnamefont
  {Oliveira}},\ }\bibfield  {title} {\enquote {\bibinfo {title} {An exact
  solution for tube and slit flow of a {FENE-P} fluid},}\ }\href@noop {}
  {\bibfield  {journal} {\bibinfo  {journal} {Acta Mechanica}\ }\textbf
  {\bibinfo {volume} {158}},\ \bibinfo {pages} {157--167} (\bibinfo {year}
  {2002})}\BibitemShut {NoStop}%
\bibitem [{\citenamefont {Bird}, \citenamefont {Armstrong},\ and\ \citenamefont
  {Hassager}(1987)}]{bird1987dynamics}%
  \BibitemOpen
  \bibfield  {author} {\bibinfo {author} {\bibfnamefont {R.~B.}\ \bibnamefont
  {Bird}}, \bibinfo {author} {\bibfnamefont {R.~C.}\ \bibnamefont
  {Armstrong}},\ and\ \bibinfo {author} {\bibfnamefont {O.}~\bibnamefont
  {Hassager}},\ }\href@noop {} {\emph {\bibinfo {title} {Dynamics of polymeric
  liquids. Vol. 1: Fluid mechanics}}}\ (\bibinfo  {publisher} {John Wiley},\
  \bibinfo {year} {1987})\BibitemShut {NoStop}%
\bibitem [{\citenamefont {Priymak}\ and\ \citenamefont
  {Miyazaki}(1998)}]{priymak1998accurate}%
  \BibitemOpen
  \bibfield  {author} {\bibinfo {author} {\bibfnamefont {V.}~\bibnamefont
  {Priymak}}\ and\ \bibinfo {author} {\bibfnamefont {T.}~\bibnamefont
  {Miyazaki}},\ }\bibfield  {title} {\enquote {\bibinfo {title} {Accurate
  navier--stokes investigation of transitional and turbulent flows in a
  circular pipe},}\ }\href@noop {} {\bibfield  {journal} {\bibinfo  {journal}
  {Journal of Computational Physics}\ }\textbf {\bibinfo {volume} {142}},\
  \bibinfo {pages} {370--411} (\bibinfo {year} {1998})}\BibitemShut {NoStop}%
\bibitem [{\citenamefont {Meseguer}\ and\ \citenamefont
  {Trefethen}(2003)}]{meseguer2003linearized}%
  \BibitemOpen
  \bibfield  {author} {\bibinfo {author} {\bibfnamefont {A.}~\bibnamefont
  {Meseguer}}\ and\ \bibinfo {author} {\bibfnamefont {L.~N.}\ \bibnamefont
  {Trefethen}},\ }\bibfield  {title} {\enquote {\bibinfo {title} {Linearized
  pipe flow to reynolds number $10^7$},}\ }\href@noop {} {\bibfield  {journal}
  {\bibinfo  {journal} {Journal of Computational Physics}\ }\textbf {\bibinfo
  {volume} {186}},\ \bibinfo {pages} {178--197} (\bibinfo {year}
  {2003})}\BibitemShut {NoStop}%
\bibitem [{\citenamefont {Malik}\ and\ \citenamefont
  {Skote}(2019)}]{malik2019linear}%
  \BibitemOpen
  \bibfield  {author} {\bibinfo {author} {\bibfnamefont {M.}~\bibnamefont
  {Malik}}\ and\ \bibinfo {author} {\bibfnamefont {M.}~\bibnamefont {Skote}},\
  }\bibfield  {title} {\enquote {\bibinfo {title} {A linear system for pipe
  flow stability analysis allowing for boundary condition modifications},}\
  }\href@noop {} {\bibfield  {journal} {\bibinfo  {journal} {Computers \&
  Fluids}\ }\textbf {\bibinfo {volume} {192}},\ \bibinfo {pages} {104267}
  (\bibinfo {year} {2019})}\BibitemShut {NoStop}%
\end{thebibliography}%
\end{document}